\input epsf

\documentstyle[times, emulateapj]{article}

\def\be{\begin{equation}}
\def\ee{\end{equation}}
\def\bea{\begin{eqnarray}}
\def\eea{\end{eqnarray}}

\def\spose#1{\hbox to 0pt{#1\hss}}

\newcommand\lessim{\mathrel{\spose{\lower 3pt\hbox{$\mathchar"218$}}
     \raise 2.0pt\hbox{$\mathchar"13C$}}}
\newcommand\gtsim{\mathrel{\spose{\lower 3pt\hbox{$\mathchar"218$}}
     \raise 2.0pt\hbox{$\mathchar"13E$}}}
\def\cyg{\hbox{Cyg X-1}}
\def\gx{\hbox{GX 339--4}}

\begin{document}

\slugcomment{To appear in The Astrophysical Journal, Letters}

\lefthead{Vaughan \& Nowak}
\righthead{X-ray Variability Coherence: How to Compute it, What it Means,
$\ldots$}

\title{ X-ray Variability Coherence: How to Compute it, What it Means,
and How it Constrains Models of GX339--4 and Cygnus X--1}

\author{Brian A. Vaughan$~^*$ and Michael A. Nowak$~^\dagger$} 
\affil{$^*$ MC 220-47, Caltech, Pasadena, CA ~91125; ~brian@srl.caltech.edu}
\affil{$^\dagger$ JILA, University of Colorado, Campus Box 440, Boulder, CO
~80309-0440; ~mnowak@rocinante.colorado.edu}

\affil{\sl Received September 6, 1996; Accepted September 30, 1996}
\received{September 6, 1996}
\accepted{September 30, 1996}

\begin{abstract}
We describe how the coherence function---a Fourier frequency-dependent
measure of the linear correlation between time series measured
simultaneously in two energy channels---can be used in conjunction with
energy spectra, power spectra, and time delays between energy channels to
constrain models of the spectrum and variability of X-ray binaries.  Here
we present a procedure for estimating the coherence function in the
presence of counting noise.  We apply this method to the black hole
candidates Cyg X--1 and GX 339--4, and find that the near perfect coherence
between low- and high-energy X-ray photons rules out a wide range of models
that postulate spatially extended fluctuating emission, thermal flares,
and overlapping shot-noise.
\end{abstract}

\keywords{black hole physics --- methods: statistical --- X--rays: stars}

\section{Introduction}
Rapid ($T<1000\,$s) aperiodic variability is common to all types of X-ray
binaries.  It arises in the immediate vicinity of the compact object and
provides a probe of changes in physical parameters such as accretion rate,
optical depth, and temperature.  Power spectra and cross spectra have been
applied widely to study variability in X-ray binaries as a function of
frequency and energy, as well as to measure time delays between intensity
variations at different energies (cf. van der Klis 1989).  Here we describe
the coherence function---a measure of the degree of linear correlation
between two time series as a function of Fourier frequency---and how this
provides strong constraints on models of X-ray binaries, especially recent
theoretical models that attempt to correlate energy spectra with aperiodic
variability behavior (cf. Nowak 1994, Miyamoto et al. 1994, Miller 1995,
Nowak \& Vaughan 1996).  We expect the coherence function to be widely
applicable to {\it Rossi X-ray Timing Explorer (RXTE)} observations.

No current model mimics all facets of a system as complex as an X-ray
binary.  However, spectral and dynamical models should qualitatively
reproduce --- or at the very least not fundamentally disagree with --- {\it
all} of their most basic average properties: the energy spectrum, power
spectrum (PSD), and time delays and coherence between different energies.
(For ways in which a {\it spectral} model can disagree with {\it timing}
data, cf. Miller 1995, Nowak \& Vaughan 1996.)  Most theories model the
energy spectra, fewer model the PSD, fewer still consider phase lags, and,
to date, none consider the coherence function.  Here we describe several
examples that lead to unity coherence, as well as several that generically
lead to a loss of coherence.

We apply the coherence function to the black hole candidate \cyg~ in its
low (hard) state and \gx~ in its very high state.  The former shows unity
coherence over a wide range of Fourier frequencies and energy bands,
whereas the latter shows a sharp drop in coherence between low- and
high-energy bands.  We present these data to show the application of the
statistical methods to real data, and to show that there are intrinsic
differences among distinct physical states and systems.

\section{Computing the Coherence Function}

Let $x_1(k)$ and $x_2(k)$ be the X-ray flux observed simultaneously in two
energy channels at times $t_k$.  The Fourier transforms of $x_1(k)$ and
$x_2(k)$ at frequency $\nu_j$ are $X_1(j)$ and $X_2(j)$, respectively.  We
will usually use lowercase variables to indicate time series and uppercase
variables to indicate Fourier transforms.  From the Fourier transforms one
can construct the power spectra [$P_1(j)=|X_1(j)|^2$ and
$P_2(j)=|X_2(j)|^2$] and the (complex-valued) cross spectrum
[$C(j)=X_1^*(j)X_2(j)$], whose argument is the phase difference between
intensity fluctuations in the two channels at frequency $\nu_j$ which can
be converted to a time delay $\delta t(j) =
\arg[C(j)]/(2\pi\nu_j)$.

Measured X-ray light curves always contain noise.  For most X-ray
observations, the dominant noise source is counting, or Poisson, noise.  We
denote signal by $s$ and noise by $n$, and write $x_1(k) = s_1(k) +
n_1(k)$, $X_1(j) = S_1(j) + N_1(j)$, and likewise for channel 2.  Hereafter
we usually drop the explicit frequency dependence to simplify notation.
For power spectra, Poisson noise adds to the signal, and $P_1 = |S_1|^2 +
|N_1|^2$.  The signal power, $|S_1|^2$ can be estimated by subtracting the
Poisson noise, $|N_1|^2$, from the measured power.  

We now consider the idealized case of two simultaneous random processes, 1 and
2, and let $s_1(t)$ and $s_2(t)$ be noiseless signals drawn from these
processes.  It is usually possible to find a linear transform, $h(\tau)$,
called the transfer function, that relates $s_1(t)$ and $s_2(t)$ via
\be
s_2(t)~=~\int_{-\infty}^{\infty}h(t-\tau)s_1(\tau)~d\tau,
\label{time_transfer}
\ee
or, equivalently, $S_2(f)=H(f)S_1(f)$, where $S$ and $H$ are the Fourier
transforms of $s$ and $h$.  If $H(f)$ is the same for all
realizations of the two processes, the processes are said to be
coherent at frequency $f$.  In that case, 
$|\left<C(f)\right>|^2 =$ $|H(f)|^2\left<|S_1(f)|^2\right>^2=$
$\left<|S_1(f)|^2\right>\left<|S_2(f)|^2\right>$,
where angle brackets denote an average over an infinite set of signals from the
same processes. The coherence function is then defined as
\be
\gamma_I^2(f)~\equiv~{{|\left<C(f)\right>|^2} \over 
     {\left<|S_1(f)|^2\right>\left<|S_2(f)|^2\right>}},
\label{gamma_I^2(f)}
\ee
where we affix a subscript $I$ when computing the intrinsic coherence
between noiseless signals.  It is clear that when
$S_1(f)$ and $S_2(f)$ are related by a linear transform, $\gamma_I^2(f)=1$.
Coherence is a measure of the fraction of the rms amplitude of one process
at a given frequency that can be predicted from the other by a linear
transform (Bendat \& Piersol 1986).  It is important to note that unlike
powers and time delays, coherence can only be computed for an ensemble of
independent measurements.  For noiseless measurements from a finite number,
$m$, of independent samples, the statistical uncertainty of the coherence
function is $\delta \gamma_I = (2/m)^{1/2} ~(1-\gamma^2_I)/|\gamma_I|$
(Bendat \& Piersol 1986).

The coherence function of real data is given by analogy to
equation~(\ref{gamma_I^2(f)}), except that angle brackets are used to
denote an average over a finite number, $m$, of independent measurements.
The coherence function of noisy data will always be less than unity.  The
intrinsic coherence can be estimated by correcting each term in the
discrete analogy to equation~(\ref{gamma_I^2(f)}) for counting noise.  For
the powers in the denominator, one sets $|S_1|^2=P_1-|N_1|^2$, and
similarly for $|S_2|$, where $P_1$ and $P_2$ are the measured, noisy power
spectra of the signals.

Writing out $|\left<C\right>|^2$ in terms of signal and noise yields 
\be
|\left<C\right>|^2~=~|\left<S_1^*S_2\right>+\left<S_1^*N_2\right>
+\left<N_1^*S_2\right>+\left<N_1^*N_2\right>|^2~~.
\label{long_C}
\ee
The first term is the average cross spectrum of the signals.  Poisson noise
in one channel is uncorrelated with Poisson noise in the other channel and
with signal in either channel.  Thus, $|\left<C\right>|^2$ is the squared
magnitude of the sum of $S_1^*S_2$ and a complex random variable, $\varsigma$,
whose real and imaginary parts each have zero mean and variance
\be
{\rm var}(\varsigma)~=~ {1 \over 2} (|S_1|^2|N_2|^2+|N_1|^2|S_2|^2
     +|N_1|^2|N_2|^2)/m~~.
\label{variance}
\ee
We define $s^2 \equiv |\left<S_1^*S_2\right>|^2$, $a^2 \equiv
|\left<C\right>|^2$, and 
$n^2 \equiv (|S_1|^2|N_2|^2+|N_1|^2|S_2|^2+|N_1|^2|N_2|^2)/m$. 
The probability distribution of the measured cross spectrum, 
$|\left<C\right>|^2$,  given an intrinsic cross spectrum and Poisson noise,
is equivalent to the distribution of total power in an individual frequency
bin of a noisy power spectrum; it is given by
\be
p(a^2|s^2,n^2)~=~{n^{-2}}\exp\left[{{-(a^2+s^2)} \over {n^2}}\right]
               I_0\left({{2as} \over {n^2}}\right)
\label{p(a)}
\ee
(Groth 1975, Goodman 1985), where $I_0$ is the zero order Bessel function of
the first kind.  This can be inverted to yield the probability distribution
of $s^2$ given a measured cross spectrum and Poisson noise,
\be
p(s^2|a^2,n^2)={{(ns)^{-2}}\over{\sqrt{\pi}}}
                \exp\left[{{-(a^2+2s^2)} \over {2n^2}}\right]
{{I_0({{2as}/{n^2}})}\over{{I_0({{a^2}/{2n^2}})}}} .
\label{p(s2)}
\ee
(Chakrabarty 1996).  For $a\gg n$, equation~\ref{p(a)} can be 
approximated by the Gaussian
\be
p(a^2|s^2,n^2) ~\approx~ {1 \over {2na\sqrt{\pi}}}
\exp\left[{{-(a-s)^2} \over {n^2}}\right],
\label{approx-p(s)}
\ee
in which case $\left<a^2\right>=\left<s^2\right>+\left<n^2\right>$ and
$\delta a^2=\sqrt{2}n^2$.  
The intrinsic coherence can be usefully estimated when the
following conditions are met: $s\gtsim n$,
$|S_1|^2/|N_1|^2\gtsim1/\sqrt{m}$, and $|S_2|^2/|N_2|^2\gtsim1/\sqrt{m}$.

The following is a recipe for estimating the value and uncertainty of
the intrinsic coherence from measurements of $\left<C\right>$, $P_1$ and $P_2$.
Average powers and cross spectra should be constructed from unnormalized
measurements.  In practice it will often be necessary to correct the 
Poisson noise level for instrumental dead time (cf. van der Klis 1989 for
a detailed discussion).  Dead time can introduce correlations between
energy channels that enhance or diminish coherence.  These are intrument
dependent and beyond the scope of this Letter.  The terms high power
and high measured coherence used below denote powers that satisy 
$|S|^2$ greater than a few times $|N|^2/\sqrt{m}$ 
in each channel and measured coherence that satisfies
$\gamma^2$ greater than a few times $n^2/(P_1P_2)$, respectively.
In many cases, these conditions will be
satisfied at some frequencies but not at others.

\smallskip
{\sl High powers, high measured coherence.---} 
\bea
\gamma_I^2 &=& { {|\left<C\right>|^2-n^2} \over {|S_1|^2|S_2|^2} }
     \Bigg ( 1~\pm~ m^{-1/2} \Bigg [~ { {2n^4 m} \over 
     {(|\left<C\right>|^2-n^2)^2 } }
     \nonumber \\
     &+& { {|N_1|^4} \over {|S_1|^4} } + { {|N_2|^4} \over {|S_2|^4} } +
     { {m \delta \gamma_I^2} \over {\gamma_I^4} } ~ \Bigg ]^{1/2} ~ \Bigg )
     ~~.
\label{gamma-high}
\eea
This is the Gaussian limit, and in essence it corresponds to optimally
filtering the measured coherence.  The first three terms in the uncertainty
come from uncertainties in Poisson noise, and the last term is from the
statistical uncertainty in the intrinsic coherence.

\smallskip
{\sl High powers, low measured coherence.---}Use 
equation~(\ref{p(s2)}) to determine
confidence limits, $|\left<S_1^*S_2\right>|^2_{\rm min}$ and 
$|\left<S_1^*S_2\right>|^2_{\rm max}$,
on $|\left<S_1^*S_2\right>|^2$.  Confidence limits on $\gamma^2_I$ are
then 
\be
\gamma^2_{I,{{\rm max} \atop {\rm min}}} = 
     { { |\left<S_1^*S_2\right>|^2_{{\rm max} \atop {\rm min}} }
     \over  { |S_1|^2 |S_2|^2 } }
     \left ( 1 \pm \sqrt{ { {|N_1|^4} \over {m|S_1|^4} } 
     + { {|N_2|^4} \over {m|S_2|^4} }  
     + { { \delta \gamma_I^2 } \over {\gamma_I^4} } }  \right ) ~.
\ee

\smallskip
{\sl Low powers.---}This difficult case
arises in weak sources and at high frequency.  In practice the $1\sigma$
errors in this case are likely to extend nearly from 0 to 1.  The authors
do not know a closed form for $p(\gamma_I^2)$.  We recommend using
equation~(\ref{p(s2)}) and the Gaussian probability distributions of 
$|S_1|^2$ and $|S_2|^2$ to emperically map out $p(\gamma_I^2)$ and 
determine confidence limits on the intrinsic coherence.

\section{Examples of Incoherent Sources}

\smallskip
\noindent{\sl Thermal Flares.---} Let us imagine that the observed variability
is completely the result of local temperature fluctuations with some time
dependence, $T(t)$, which can be arbitrarily complicated so long as it is
statistically stationary.  Let the intrinsic time series, $s_1(t)$ and
$s_2(t)$, be photon count rates in narrow frequency bands $(\nu_1, \nu_1 +
\delta \nu_1)$ and $(\nu_2, \nu_2 + \delta \nu_2)$.  The observed count
rate in frequency band 1 is approximately proportional to $\nu_1^2 \delta
\nu_1 \left ( \exp \left [ {{h \nu_1}/{kT(t)}} \right ] ~-~ 1 \right )^{-1}
$, and likewise for energy band 2.  If both $h\nu_1$ and $h \nu_2 \ll k
T(t)$, then we are on the Rayleigh-Jeans portion of the spectrum, in which 
case $s_1(t) \propto s_2(t) \propto T(t)$.  Thus there is a linear transfer
function (a constant) between the two channels, and therefore the coherence
function will be unity. On the other hand, if $kT(t) \ll h \nu_1$ and $h
\nu_2$, then we are on the Wien tail of the spectrum, in which case we have
$s_2(t) \propto s_1(t)^{(\nu_2/\nu_1)} \propto \exp[-h \nu_2/kT(t)]$.  
Thus there is a nonlinear transfer function.  In general, such a transfer 
function will take power from a frequency $f$ and distribute it among 
harmonics of $f$ (cf. Bendat \& Piersol 1986) and therefore lead to a loss of 
coherence.

\smallskip
\noindent{\sl Multiple Flaring Regions.---}A number of models for X-ray
variability in Galactic black hole candidates associate different
timescales with different physical regions of an accretion disk (cf. Nowak
1994, Miyamoto et al. 1994).  If more than one region contributes to the
signal in both energy bands, then it is possible for the coherence function
to be less than unity, even if {\it individual} regions produce
perfectly coherent variability.

Consider two flaring regions: one produces a time series $q(t) = q_1(t) +
q_2(t)$, and the other produces a time series $r(t) = r_1(t) +
r_2(t)$---each in energy bands 1 and 2, respectively.  Furthermore, 
assume that there
is a constant linear transfer function that relates $q_1(t)$ to $q_2(t)$,
as well as another constant, linear transfer function that relates $r_1(t)$ to
$r_2(t)$, with $q_1(t)$ and $r_1(t)$ otherwise being completely uncorrelated.
Denoting the intrinsic time series observed in band 1 as
$s_1(t) = q_1(t) + r_1(t)$, and the intrinsic time series observed in band
2 as $s_2(t) = q_2(t) + r_2(t)$, the intrinsic coherence function between
the two bands is then 
\be
\gamma^2_I = {{ Q_1^2 Q_2^2 + R_1^2 R_2^2 + 2 
     | Q_1 | | Q_2 | | R_1 | | R_2 |  \cos ( \delta \theta_r - \delta
     \theta_q  ) } \over { Q_1^2 Q_2^2 + R_1^2 R_2^2 + Q_1^2 R_2^2 + 
     Q_2^2 R_1^2 }} ,
\label{multi_coh}
\ee
where $\delta \theta_q$ and $\delta \theta_r$ are the mean Fourier phase
differences between $Q_1(f)$ and $Q_2(f)$ and between $R_1(f)$ and
$R_2(f)$, respectively.   (The quantities on the right hand side of 
equation (\ref{multi_coh}) refer to their mean values.)
One can easily show that for this case $\gamma^2_I \le 1$, and is only
equal to unity if both $\delta \theta_q = \delta \theta_r$ and $| Q_1 |/|
Q_2 | = | R_1 |/| R_2 |$.  That is to say, the coherence function is unity 
{\it if and only if} the same linear transfer function that takes $q_1(t)$ 
to $q_2(t)$ also takes $r_1(t)$ to $r_2(t)$.

\section{Examples of Coherent Sources}

Here we present an idealized model that produces unity coherence while
producing phase lags qualitatively similar to those seen in Cyg X--1
(i.e. phase lags approximately independent of Fourier frequency; Miyamoto
et al. 1992).  Imagine that we have a disk with a stochastic source of 
(linear) surface density perturbations at the center that propagate
outwards at some constant speed $c_s$.  If we take the source to be given
by $(4 \pi r)^{-1} \delta(r) \sigma_s(t)$, then the resultant surface
density perturbations---$\sigma_p(r, t)$, at radius $r$ and time $t$ 
(using the two-dimensional wave propagation Green's function; cf. Morse \& 
Feschbach 1953)---are given by
\be
\sigma_p (r,t) ~=~ \int dt' ~\sigma_s(t') ~ {{\theta[~ (t-t') ~-~ r/c_s ~]}
 \over {\sqrt{~ (t-t')^2 ~-~ r^2/c_s^2 ~} } } ~~,
\label{two_d}
\ee
where $\theta$ is the step function.   Let the fluctuating signal in 
band 1, $s_1(t)$, be equal to the surface density perturbation times 
a weighting function, $g_1(r)$, integrated over the
entire disk.  The radially integrated Green's function is just a linear
transfer function, thus the Fourier transform of $s_1(t)$ becomes
\be
S_1(f) ~=~ \Sigma_s(f) ~ \int i 2 \pi^2 ~r g_1(r) ~ H_0^{(1)} \left (
     {{2 \pi |f| r}\over{c_s}} \right ) ~dr 
     ~~,
\label{two_dtrans}
\ee 
which we define as $\Sigma_s (f) T_1 (f)$, and where
$H_0^{(1)}(x)$ is the Hankel function of $x$ as well as the Fourier transform
of the two-dimensional Green's function.  Similarly we have 
$S_2(f)\equiv \Sigma_s(f) T_2 (f)$.  As both $S_1$ and $S_2$ are linearly 
related to the source $\Sigma_s$, there is a linear transfer function 
equal to $T_2/T_1$ that takes $S_1$ to $S_2$, and therefore there is unity 
coherence.  We could write such a transfer function only because the
driving source was separable in time and space (in this case being 
spatially localized).

Equation (\ref{two_dtrans}) is analytically tractable if we set $g_1(r) =
\exp(-\alpha_1 r/r_0 )$ and $g_2(r) = \exp(-\alpha_2 r/r_0)$, as one might
have for weighting functions representing flux from a Wien tail.  As will
be discussed elsewhere (Vaughan et al. 1997), the resulting time lags are
qualitatively similar to those seen in Cyg X--1.  We require $r_0/c_s
\sim 1$ to obtain quantitative agreement with the observations, which
implies an extremely slow propagation speed, $c_s \sim 10^{-4} c$.  This
mechanism is qualitatively similar to that proposed by Manmoto et
al. (1996), where thermal waves are launched from large disk radii toward
the center, and then reflect from the disk inner edge as acoustic waves.
However, the mechanism of Manmoto et al. (1996) cannot preserve coherence
if, as in the example of \S3, there are multiple, spatially distributed
source functions ({\it in the X-ray-emitting regions}) for the waves.
Furthermore, the acoustic waves shock upon reflection, which is an
inherently non-linear process that also will lead to a loss of coherence,
if emission from the reflected shock is an appreciable fraction of the
total observed emission.

Adding additional mechanisms in the form of successive transfer functions
will lead to unity coherence so long as each component is itself linear.
In the above example, we could have invoked Comptonization as a subsequent
transfer mechanism, which would have introduced additional time delays.  So
long as the Compton cloud is static, coherence will be preserved (cf. Nowak
\& Vaughan 1996).

\section{Observational Examples}

Of the above physical situations where coherence is preserved and those
where it is not, which cases are observed in nature?  Surprisingly, the
answer seems to be that unity coherence is the norm, despite the variety of
ways of weakening coherence.  Here we present two black hole candidates:
GX339--4 in its very high state, and Cyg X--1 in its low state.  Both
have been modeled with disks plus Compton coronae (cf. Miyamoto et
al. 1991; Dove, Wilms, \& Begelman 1996 and references therein), although
there are differences between them:  the very high state has a substantial
soft ($\sim 1$ keV) component, the low state has none;  the hard tail of
the very high state is softer than the hard tail of the low state ($\sim
2.5$ photon index compared to $\sim 1.7$);  the very high state has lower
variability (rms $\sim 10-20\%$) than the low state (rms $\sim 40\%$).
Yet, as shown in Figure 1, both systems apparently have unity intrinsic
coherence, at least between low energies and at low frequency.

Figure 1 presents coherence plots derived from {\it Ginga} data.  
The methods described above have been used to
filter the noise from the data. Poisson noise and uncertainties in the
{\it Ginga} dead time make the coherence estimates unreliable 
above $\sim 10$ Hz ; however, at frequencies below 10~Hz the
intrinsic coherence is essentially unity.  This result holds true for Cyg
X--1 over a wide range of energy channels.  On the other hand, the GX 339--4
variability (the PSD is shown as Fig. {\it 4b} in Miyamoto et al. 1991) shows
unity coherence between low energies but shows a dramatic drop in
coherence between the $2.3-4.6$ keV and $13.9-37.1$ keV channels.

Considering the mechanisms discussed above, this loss of coherence is not
surprising. Perhaps the $13.9-37.1$ keV band corresponds to a Wien tale,
and, therefore, to a nonlinear transfer function. It is more difficult to
explain the observed unity coherence.  We expect that most shot models
(cf. Lochner et al. 1991) will produce less than unity coherence.  For
example, the kinematic model of Nowak (1994) for the very high state of GX
339--4 successfully reproduced the observed PSDs and phase lags. However,
as the viscous and thermal fluctuations were distributed over a large range
of radii and overlapped in time, the mechanism led to less than unity
coherence (not shown in that paper) between {\it all} energy bands.  In
order to preserve coherence, we always require linear responses, and we
usually require the following: localized sources and/or localized
responses; a (temporally) uniform source throughout the disk; or a uniform
response throughout the disk.  Most of these features are absent from
current models.

The data described above will be presented in greater detail in Vaughan et
al. (1997).  Here we have presented them to demonstrate that there are
observational and, by extension, physical differences between the systems,
despite the fact that they are both commonly fit with coronae models.  The
coherence function therefore offers us another tool to help distinguish
among theoretical models.

\section{Conclusions}

We have described a statistic, the coherence function, that is derivable
from the cross spectrum used to compute phase or time lags between two
times series.  This statistic has traditionally been ignored; however, it
contains additional information about the system, and the methods described
above can minimize the noise effects.  Whenever one calculates the phase
lag, one can and should also calculate the coherence function.

In general, coherence will be lost whenever there is a nonlinear transfer
function between two channels, or whenever there are multiple, uncorrelated
(linear or nonlinear) transfer functions between two channels.  However, our
experience has been that coherence is preserved more often than
not.  This is a challenge for theoretical models, as there are many more
mechanisms for destroying coherence than there are for preserving
coherence.

We presented data from two black hole candidates: GX 339--4 in its very high
state and Cyg X--1 in its low state.  At low frequency and low energy both
preserve coherence, although GX 339--4 loses coherence between high- and
low-energy bands.  Theoretical models have addressed observations of phase
and time lags (cf. Nowak 1994, Manmoto et al. 1996, etc.), as measured with
{\it EXOSAT}, {\it Ginga}, and other X-ray timing instruments. No model 
currently accounts for the coherence properties detected in GX 339--4 and 
Cyg X--1.  Forthcoming {\it RXTE} observations, however, coupled with the 
tools presented above, can help provide new insights into the physical 
mechanisms at work in these systems, as well as new challenges for
theorists to meet.

\acknowledgements
The authors would like to acknowledge useful conversations with Peter
Michelson, Michiel van der Klis, Mitch Begelman, James Dove, and J\"orn
Wilms, as well as generous support from NASA grants NAG 5-3239  (B.A.V.),
NAG 5-3225 (M.A.N.), and the Netherlands Organization for Scientific
Research (NWO) under grant PGS 78-277.


\begin{figure}

\centerline{\epsfxsize=0.65\hsize  {\epsfbox{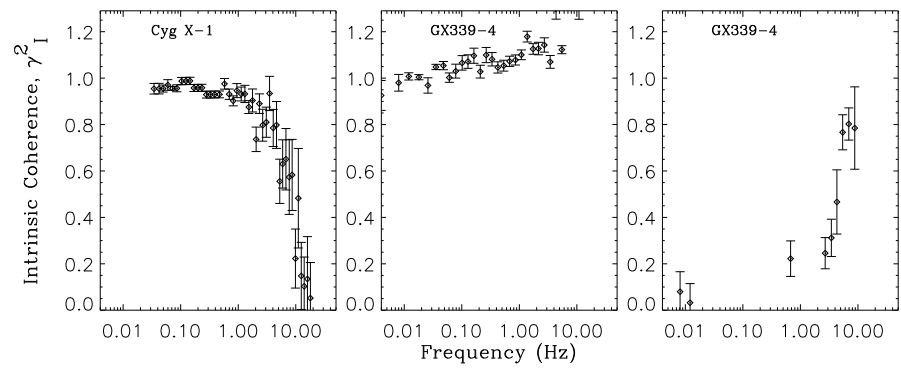}}}
\vskip -4.5 true in

\figcaption
{Coherence plots derived from {\it Ginga} data of Cyg X--1 in its low 
state ({\sl left:} $1-2.3$ vs. $9.2-11.5$ keV), and GX 339--4 in its 
very high state ({\sl middle:} $2.3-4.6$ vs. $4.6-9.2$ keV; {\sl right:} 
$2.3-4.6$ vs. $13.7-37.1$ keV). The high-frequency rollover in Cyg X--1 
is likely due to a misestimation of the Poisson noise level.  Coherence 
exceeding unity in GX 339--4 may be the result of dead time effects.  
The rapid coherence rise on the right is at the quasi-periodic oscillation 
frequency (cf. Miyamoto et al. 1991).}

\end{figure}

\end{document}